\documentclass[12pt,preprint]{aastex}

\newcommand{\ang}{$\mbox{\AA}$}
\newcommand{\sol}{$_{\odot}$}
\newcommand{\super}[1]{$^{#1}$}

\def\HST{{\it HST}\/}

\def\kms{km~s$^{-1}$}

\def\gtsim{
\mathrel{\raise.3ex\hbox{$>$}\mkern-14mu\lower0.6ex\hbox{$\sim$}}
}
\def\ltsim{
\mathrel{\raise.3ex\hbox{$<$}\mkern-14mu\lower0.6ex\hbox{$\sim$}}
}
\def\farcs{\hbox{$.\!\!^{\prime\prime}$}}

\shorttitle{Structure and Age of 1E~0102--7219}
\shortauthors{Finkelstein et al.}

\begin{document}
\title{Optical Structure and Proper-Motion Age of the Oxygen-rich Supernova Remnant
1E~0102--7219 in the Small Magellanic Cloud\altaffilmark{1}}

\author{Steven L. Finkelstein\altaffilmark{2}, Jon A. Morse\altaffilmark{2}, 
James C. Green\altaffilmark{3}, Jeffrey L. Linsky\altaffilmark{3}, 
J. Michael Shull\altaffilmark{3}, Theodore P. Snow\altaffilmark{3}, 
John T. Stocke\altaffilmark{3}, Kenneth R. Brownsberger\altaffilmark{4},
Dennis C. Ebbets\altaffilmark{4}, Erik Wilkinson\altaffilmark{4},
Sara R. Heap\altaffilmark{5}, Claus Leitherer\altaffilmark{6}, 
Blair D. Savage\altaffilmark{7}, Oswald H. Siegmund\altaffilmark{8}, 
\& Alan Stern\altaffilmark{9}}

\affil{$^{1}$ Based on observations with the NASA/ESA Hubble Space Telescope, obtained at
the Space Telescope Science Institute, which is operated by the Association of
Universities for Research in Astronomy, Inc.\ under NASA contract No.\ NAS5-26555.}
\affil{$^{2}$ Department of Physics and Astronomy, Arizona State University,
 Tempe, AZ 85281}
\affil{$^{3}$ Center for Astrophysics and Space Astronomy, University of Colorado, Boulder, CO 80309}
\affil{$^{4}$ Ball Aerospace \& Technologies Corp., Boulder, CO 80306}
\affil{$^{5}$ NASA/Goddard Space Flight Center, Greenbelt, MD 20771}
\affil{$^{6}$ Space Telescope Science Institute, Baltimore, MD 21218}
\affil{$^{7}$ Astronomy Department, University of Wisconsin, Madison, WI 53706}
\affil{$^{8}$ Space Sciences Laboratory, University of California - Berkeley, Berkeley, CA 94720}
\affil{$^{9}$ Southwest Research Institute, Boulder, CO 80302}

\begin{abstract}

We present new optical emission-line images of the young supernova
remnant (SNR) 1E~0102--7219  (hereafter E0102)  in the Small Magellanic Cloud
obtained with the {\it Hubble Space Telescope} ({\it HST})
Advanced Camera for Surveys (ACS).  E0102 is  a member  of the  oxygen-rich 
class of SNRs showing strong oxygen, neon, and other metal-line emissions 
in its optical and X-ray spectra, and an absence of hydrogen and helium.
The progenitor of E0102 may have been a Wolf-Rayet star that underwent 
considerable mass loss prior to exploding as a Type Ib/c or IIL/b supernova.  The ejecta 
in this SNR are generally fast-moving (V $>$ 1000 \kms) and emit as they are
compressed and heated in the reverse shock. In 2003, we obtained optical 
[O~III], H$\alpha$, and continuum images with the 
ACS Wide Field Camera. The [O~III] image through the F475W filter captures the full
velocity range of the ejecta, and shows considerable high-velocity
emission projected in the middle of the SNR that was Doppler-shifted
out of the narrow F502N bandpass of a previous Wide Field and Planetary Camera 2
image from 1995.  Using these two  epochs separated by $\sim$8.5  years,
we measure the transverse expansion of the ejecta around the outer rim in
this  SNR  for the  first time  at  visible wavelengths.  From proper-motion  
measurements of 12  ejecta filaments, we estimate a mean expansion velocity for the bright 
ejecta of $\sim$2000 \kms\ and an inferred kinematic age for the SNR of 
$\sim2050 \pm 600$ years.  The age we derive from {\it HST} data 
is about twice that inferred by Hughes et al.\ (2000, ApJ, 543, L61)
from X-ray data, though our 1-$\sigma$ error bars overlap. Our proper-motion age
is consistent with an independent optical kinematic age 
derived by Eriksen et al.\ (2003, AIPC, 565, 419)
using spatially resolved [O~III] radial-velocity data.  We derive an expansion center that 
lies very close to conspicuous X-ray and radio hotspots, which could indicate the presence 
of a compact remnant (neutron star or black hole).    

\end{abstract}

\keywords{ISM: supernova remnants: individual (1E~0102--7219) --- stars: Wolf-Rayet}

\section{Introduction} \label{introduction}

Supernova remnants (SNRs) reveal how stars synthesize  and
distribute the elements heavier than H and He.  Young SNRs with
uncontaminated debris are among our best natural laboratories for 
understanding how nucleosynthesis operates in massive stars.  Studying
the SNR ejecta provides clues about the dynamics of supernova
(SN) explosions, the composition and spectral type of progenitors,
and ultimately about the chemical enrichment process in galaxies.

The oxygen-rich SNR 1E~0102--7219 (E0102) in the Small Magellanic
Cloud (SMC) offers an exceptional opportunity for multi-wavelength, high
resolution studies of a young SNR. E0102 is a vital object in the
small class of oxygen-rich SNRs that contain filaments of
uncontaminated debris from core-collapse supernovae.  Out of the few
($\sim$10) known objects in the oxygen-rich class, E0102 is one of the
most interesting because of its similarities to the prototype O-rich
SNR, Cassiopeia~A (Cas~A) in the Galaxy.  However, unlike Cas~A,
the sight-line to E0102 in the SMC has low interstellar extinction
that allows us to gather data from the X-rays to the radio; yet it
is close enough that we can still discern physically important scales.

E0102 was discovered by \cite{seward81} in an {\it Einstein} X-ray observatory
survey of the Magellanic Clouds. It is located in the SMC just
northeast of the high-excitation H~II region N76A (\citealt{hen56}).
Dopita, Tuohy, \& Mathewson (1981) recognized E0102 as an O-rich SNR by
virtue of its strong optical [O~III]$\lambda$$\lambda$4959,5007 emission but complete lack
of hydrogen emission from a network of filaments. Subsequent spectra
by \cite{td83} confirmed the presence of fast-moving, highly processed ejecta
remarkable for showing emission from O and Ne only. They also estimated a
kinematic age for the SNR of $\sim$1000 years based on the velocity range they
observed and its angular size.

E0102 exhibits the textbook forward/reverse shock structure that forms
during the evolution of SNRs (\citealt{chev82}).
Emissions from the faint $\sim$44$^{\prime\prime}$ diameter blast wave and 
bright $\sim$30$^{\prime\prime}$ diameter shell of shocked ejecta have now been observed
in exquisite detail across the electromagnetic spectrum from the ground and
space (e.g., \citealt{amyball,blair89,hay94,hughes00,gaetz00,stan05}).
The combined X-ray, UV, and optical emission spectra of the ejecta contains lines
from O, Ne, C, Si, and Mg over a wide range of ionization states,
from neutral to H-like species (\citealt{blair00,ras}).
The comprehensive study by \cite{flan04} of the high-resolution {\it Chandra}
X-ray spectrum shows that the highly ionized species are stratified; the higher
ionization states occur at larger radii as the reverse shock propagates
into the expanding ejecta. These authors also estimated a progenitor zero-age-main-sequence
(ZAMS) mass of $\sim$32M\sol\ based on their derived oxygen mass of 6M\sol\ and heavy element
abundance ratios, and comparing to theoretical predictions.

In this paper we present new optical images from the Advanced Camera for
Surveys (ACS) aboard the {\it Hubble Space Telescope} ({\it HST}). We describe
the observations in Sec.\ 2 and present the imaging results in Sec.\ 3. The
new images capture the full velocity range of the emitting ejecta and reveal
fresh details on the fastest filaments projected against the interior of the
remnant.  We compare our new [O~III] image to one obtained previously with
the Wide Field and Planetary Camera 2 (WFPC2) and measure proper motions
of several outer filaments captured in both data sets. In Sec.\ 4 we
discuss some implications of our proper-motion measurements and examine
additional clues about the possible progenitor of E0102. Our conclusions
are presented in Sec.\ 5. We assume a distance to the SMC of 59 kpc (e.g.,
van den Bergh 1999), where $0.1'' \approx 9 \times 10^{16}$ cm.

\section{Observations} \label{observations}

\subsection{HST/ACS Wide Field Camera Images}

We obtained images of E0102 with the ACS Wide Field Camera (WFC) aboard
the {\it Hubble Space Telescope} in October 2003 (see Table~1) to
select positions for future UV spectroscopic observations with the
Cosmic Origins Spectrograph (COS; \citealt{green03}). Although a previous {\it HST} image
through the F502N filter of WFPC2 exists (GO-6052, PI Morse;
\citealt{blair00}), many of the brightest O-rich filaments through the
interior of the SNR were Doppler-shifted out of the narrow bandpass.
The new ACS~WFC images were obtained using the Sloan $g$ (F475W), $i$ (F775W),
and $z$ (F850LP) broadband filters, the F550M (line-free continuum) filter,
and the F658N (H$\alpha$ + [N~II]$\lambda\lambda6548,6583$) filter
to capture the full velocity range of the emitting SNR ejecta, map the
surrounding H~II region emission, and characterize the stellar population
along this line-of-sight through the Galaxy and SMC.
We acquired the images through each filter using the
standard line-dither observing pattern to fill in the inter-chip gap
and identify hot pixels, while obtaining two exposures at each dither
position to recognize cosmic rays. We used the IRAF/STSDAS\footnote[10]{IRAF is
distributed by the National Optical Astronomy Observatories (NOAO), which is
operated by the Association of Universities for Research in Astronomy,
Inc.\ (AURA) under cooperative agreement with the National Science Foundation.
The Space Telescope Science Data Analysis System (STSDAS) is distributed by
the Space Telescope Science Institute.} environment for most of our
image processing and analysis work. We began with the standard ACS pipeline
processed (bias-subtracted, flat-fielded, geometric distortion-corrected,
drizzled) images for our analysis. The final drizzled images had an
image scale of 0\farcs05 pixel$^{-1}$ and we measured a typical point-source
FWHM of $2.2$ pixels in the visible passband filters, corresponding to a spatial
resolution of 0\farcs11 $\approx 1 \times 10^{17}$ cm.

Figure~1 shows a labeled ground-based ESO/MAMA $U$-band image of N76A and
vicinity. The $U$-band image includes [O~II]$\lambda\lambda3727$ emission
from the N76A nebula and E0102 SNR. In Figure~1 we have drawn the footprint
of the ACS~WFC field displayed in Figures~2 and 3, centered about $0.5'$
southwest of E0102.
Our primary tracer of SNR ejecta emission is [O~III]. Referring to the E0102
optical spectra of \cite{td83} or \cite{blair00}, the ACS~WFC F475W filter
includes ejecta emission from only the [O~III] lines at $\lambda\lambda4959,5007$, 
and (weak) $\lambda4363$. Figure~2 shows an ACS color composite
of E0102 and surroundings with the F550M continuum in blue, F475W [O~III]
in green, and F658N H$\alpha$+[N~II] in red. Grayscales of the individual
[O~III] and H$\alpha$+[N~II] images are shown in separate panels of Figure~3.
Figure~4 offers a close-up view of the complex filamentary structure in
E0102, this time with the F775W image instead of F658N in red; this image
reveals significant emission toward the middle of the SNR seen for the first
time at $\sim0.1''$ spatial resolution. The bright [O~III] filaments are
detected $\sim50-60\sigma$ above the sky noise, and the fainter
structures are detected at the $\sim5-8\sigma$ level.

\subsection{Alignment with HST/WFPC2 and Chandra/ACIS Images}

Figure~5 compares the previous F502N WFPC2 Planetary Camera image and the new
F475W ACS~WFC image, each using [O~III] emission to trace the optical filaments.
The WFPC2 data were obtained $\sim$8.5 years prior to the ACS images.
We subtracted the two images in the bottom panel to highlight two features:
(1) high-velocity emission through the interior of the SNR that is now included
in the ACS bandpass (displayed in white); and (2) black-and-white edges where
we have for the first time detected proper motions of several [O~III] filaments
around the periphery of the SNR. We later use the proper motions to derive an
expansion center and age for E0102.

To detect ejecta motions between the two epochs, the [O~III] images had to be
aligned to sub-pixel accuracy.  Based on previous kinematic data
(e.g., \citealt{td83,erik01}) we expect transverse motions of only 
$\ltsim0.1''$, given our $\sim$8.5-year time baseline and 59 kpc distance to
the SMC.  We registered the 2003 ACS image to the 1995 WFPC2 image using
the IRAF tasks GEOMAP and GREGISTER in order to compensate for differences
in the image scales, rotations, and geometric distortion corrections.
We selected 40 stars in the WFPC2 image (Figure~5) to use as tie points
for registering the ACS image. The stars were chosen to provide adequate
astrometric information across the field of the SNR, as well as high enough
signal-to-noise ratios for accurate centroiding in the narrowband F502N WFPC2
Planetary Camera image without saturating in the broadband F475W ACS~WFC image.
The registration stars are circled in the left panel of Figure~6. We used
a cubic polynomial fit to transform and align the images. The fit had
r.m.s.\ residuals of $\sim$0.1 pixels, corresponding to $\sim$5 mas. Over the
8.5-year time baseline, a displacement of 5 mas corresponds to a transverse
velocity of $\sim$165 \kms\ at the distance of the SMC, which we regard as our
internal systematic uncertainty due to alignment errors. Additional uncertainties
arise during our proper-motion measurement process that we discuss in Sec.\ 3.3.
The final image scale in the registered data is that for the WFPC2 Planetary
Camera of $0.04555''$/pixel (\citealt{holtz}).

We also aligned a {\it Chandra} ACIS-S calibration image (\citealt{hughes00})
with the {\it HST} images, the same {\it Chandra} image that \cite{blair00}
used in their X-ray comparison. The {\it Chandra} image had a live-time exposure
length of 11 ksec. Since there are no common tie points (i.e., stars) in the
optical and X-ray images, we used the IRAF task WREGISTER and the spacecraft
pointing information in the image headers to align the images. The uncertainty
in the image registration is probably $\ltsim1''$ due to {\it HST} guide star
catalog astrometry errors, {\it Chandra} aspect solution errors, etc. The ACIS-S
image captures primarily soft X-ray emission in the 0.6 -- 1.5 keV range that
includes several O, Ne, and Mg emission lines (e.g., \citealt{gaetz00,hughes00}).
While there are several correspondences with the optical emission, the X-ray
emission is more complete around the periphery of the SNR, hence the {\it Chandra}
image was helpful for providing initial estimates of the SNR expansion center.
We discuss below some new correlations between the X-rays and the optical emission
seen in the ACS image, building on previous analyses
(e.g., \citealt{gaetz00,blair00}). The {\it Chandra} ACIS-S image, deconvolved with
5 iterations of the STSDAS task LUCY, is shown in the right panel of Figure~6. The
deconvolved image highlights small-scale emission features, particularly the bright
filament along the southeast limb, several spoke-like features through the middle
of the SNR, and a barely resolved blob just below the SNR geometric center.

\section{Results and Analysis} \label{results}

\subsection{Image Analysis}

In the initial E0102 SN explosion, a shock wave drove out from the surface
of the star, with the ejecta following slower behind it, but still moving
with V $>$ 1000 \kms.  As this blast wave has swept up ambient gas, it has
slowed down to velocities lower than the fastest ejecta.  A reverse
shock has thus formed that decelerates and excites the ejecta
(\citealt{blon01}). As in Cas~A, emission from ejecta that have passed
through the reverse shock in E0102 dominates its appearance in optical
and X-ray emission-line images.

The images we obtained with the ACS show much more emission than the
previous WFPC2 data.  Notwithstanding the larger field of view of the ACS~WFC
as compared with the WFPC2 Planetary Camera, the ACS filter bandpasses we
employed are also wider than those used with WFPC2. Nebular imaging with
wide bandpasses usually creates problems for understanding which emission
lines dominate in a particular region, but for young, metal-rich SNRs that
exhibit relatively sparse optical emission-line spectra the confusion is
minimized (e.g., \citealt{fesen01,morse04}). Our F475W ACS~WFC image
contains both lines of the [O~III]$\lambda\lambda4959,5007$ doublet\footnote[11]{H$\beta$ 
and higher H Balmer lines might normally be confused with [O~III] in the F475W filter, 
but the absence of H$\alpha$ emission in E0102 means that H$\beta$ is absent as well,
as shown in the spectra of \cite{td83} and \cite{blair00}.}
and easily covers the entire velocity range of $\sim6500$ \kms\ (\citealt{td83}).
However, one complication of the improved sensitivity and velocity coverage
is that many more foreground/background stars (and galaxies) appear in the
ACS images, many of which needed to be removed for the proper-motion analysis.

Figure 2 enables us to see more of the environment surrounding E0102
at high spatial resolution than previously possible. The high-ionization
H~II region N76A located in the lower-right portion of the image is highly
filamentary. It appears yellow-ish in the color composite as a blend of
primarily He~II $\lambda4686$ and [O~III]$\lambda\lambda4959,5007$ in the
green (F475W) image and H$\alpha$+[N~II]$\lambda\lambda6548,6583$ in the
red (F658N) image. The red edges of the filaments suggest that the nebula
is ionization bounded. We discuss possible implications of the proximity
of E0102 to N76A in Sec.\ 4.2.  Previous authors, including \cite{td83},
have commented on the diffuse halo of ionized gas immediately
surrounding the network of E0102 ejecta filaments. There is a $\sim$10$^{\prime\prime}$
gap between the filaments and this partial halo seen mainly in
[O~III] and partly in H$\alpha$ (Figure~3). A faint X-ray plateau and
radio emission fill this gap (e.g., \citealt{gaetz00}). The outer
edge of the X-ray plateau has been interpreted as marking the propagation of
the blast wave into the circumstellar medium and the
radio emission is likely synchrotron emission from electrons accelerated
in magnetic fields compressed behind the blast wave. \cite{hughes00}
determined that the abundances from the X-ray blast wave plateau are
consistent with shocked low-metallicity SMC interstellar gas. The optical
spectrum of the surrounding diffuse halo (\citealt{td83}) shows it to be stationary
and highly ionized, with strong [O~III] and He~II emission, reminiscent of
the extended narrow-line regions in active galactic nuclei (\citealt{morse96a}).
\cite{morse96b} showed that a similar halo around the O-rich SNR N132D in the
Large Magellanic Cloud is likely ionized by high-energy photons produced
by shocks within the SNR itself, and we favor this scenario in E0102 as well.
The H$\alpha$ + [NII] image in the bottom panel of Figure~3 shows some clumpiness
in the external medium, including a series of small cloudlets just to the
north of E0102.  These may be part of an extended plume from N76A, or they may be
wind material from the progenitor of E0102.

The close-up color composite image in Figure~4 identifies the O-rich
ejecta in green. The ejecta are composed of long sinuous filaments with
distortions and protrusions reminiscent of Rayleigh-Taylor instabilities and
individual knots down to the resolution limit of our images.
Overall the detailed structure of E0102 closely resembles the optical
filaments in Cas~A (e.g., \citealt{fesen01}), though E0102 is several
times larger in size. The optical emission is obviously asymmetrical;
the entire eastern edge is almost continuous, but the western
and northern portions show much less emission.
There is even a broken region in the southwest that
resembles the break-out of the northeastern jet in Cas~A. The difference
image in Figure~5 highlights in white the regions where we detected emission
with ACS F475W that was Doppler-shifted out of the WFPC2 F502N bandpass.
The WFPC2 F502N filter has a FWHM width of about $\pm 900$ \kms;
debris moving faster than this was therefore missed in the WFPC2 image.
The most stunning example is the bright backward S-shaped emission structure that
starts in the south-southwest, extends through the middle of the SNR, and bends
towards the northeast. In fact this is the brightest optical emission in E0102. 
The kinematic data of \cite{td83} and \cite{erik01} show that most
of the gas from this bright structure moves at about $-1000$ \kms\ at the
south and north edges and up to $-2100$ \kms\ through the middle ---
i.e., it is associated with the approaching side of the ejecta shell.
There is also bright emission from the western limb now seen in the ACS
image that likewise shows velocities of about $-1000$ \kms.

\subsection{Comparison with the X-rays}

The new ACS images afford us the chance to compare the optical emission
distribution with the X-ray emitting gas. The left image in Figure~7 shows
a color composite of the 1999 {\it Chandra} ACIS-S image with the
ACS [O~III] F475W and red continuum F775W images of E0102. The correspondences
between the X-rays and optical emission along the outer rim of E0102 have been
discussed previously by \cite{gaetz00} and \cite{blair00}. Here we are able
to add new information about the emission through the interior of the SNR.
The right panel of Figure~7 shows three color-coded contours of the
X-ray emission overlaid on the ACS F475W [O~III] image. With the full
range of [O~III] emission now available, there are several new 
interesting correlations with the X-rays. The bright southern portion 
of the backward S-shaped [O~III] complex coincides with the brightest X-ray
spoke from the southwest portion of the outer ring towards the middle of
the SNR, connecting to the central blob. The correspondence is not exact;
the spoke extends from the ring to the center in both wavelength regimes,
but in the optical the emission widens as it reaches the center. At the
point where it starts to widen, the X-ray spoke shows a bulge or knot but
then narrows to its previous thickness. About $10''$ to the east of this
spoke there is considerable optical emission with no X-ray counterparts.
The northern section of the backward S-shaped [O~III] complex likewise shows only
faint X-ray emission associated with it. However the other obvious X-ray
spoke winding from the western edge of the ring towards the middle hotspot
lies adjacent to a similar optical spoke. The X-ray spoke connects
to the central region in a small loop. Around the inner boundary of this
loop lies a similar but smaller optical loop of emission. The two spokes
with optical and X-ray emission that extend from the outer ring of emission
to the central hotspot surround an emission void perhaps best seen in the
side-by-side images of Figure~6. (The bright star in the optical image that
projects near the middle of this emission void is probably not physically
associated with the SNR.)

In summary, there are several regions through the
interior of the remnant where we see both X-ray and optical emission, but
on close scrutiny the two emissions tend to lie adjacent to each other
rather than exactly coinciding. This spatial arrangement between the optical
and X-ray emission can be understood if it represents gradients in ejecta
densities in the shocked gas.  We note that this relationship demonstrates 
the need for data from complementary wavelength regimes in order to gain a
full picture of the SNR.  Similar results are seen in Cas~A (e.g.,
\citealt{delaney04,fesen05}). We also stress that the brightest [O~III]
emission projects across the middle of the SNR, not around the edges.
The optical emission distribution implies that the ejecta are not confined
solely to a ring geometry as modeled by \cite{flan04} for the X-ray emitting
ejecta. A thick X-ray ring may still hold much of the ejecta mass but the
overall geometry of the SNR is likely more complicated. In Cas~A, for example,
the optical emission generally lies in ring-like structures situated on the
surface of an expanding shell (\citealt{law95}). Such could be the case with
much of the optical emission in E0102.

\subsection{Proper-Motion Measurements}

We used the 1995 epoch and 2003 epoch {\it HST} [O~III] images to
measure proper motions of fast-moving ejecta knots and filaments around the
periphery of E0102. Ejecta found at the largest projected radii move roughly transverse
to our line of sight and the emission is therefore captured in both the previous
narrowband F502N WFPC2 and the new broadband F475W ACS~WFC images.
The difference image in Figure~5 shows that the same features are seen
at both epochs in many locations around the SNR and that the features
have moved outward between 1995 and 2003. Measuring the transverse velocities
of fast-moving optical ejecta has long been one of our most important
tools for understanding the dynamics (e.g., expansion center and age) of Cas~A
(\citealt{mink68,vdb71,thor01,morse04,fesen05}). The exquisite spatial
resolution of {\it HST} allows us to make such measurements for the first
time in E0102 at the 59 kpc distance of the SMC. The E0102 progenitor is
thought to have had a ZAMS mass of at least 25--32 M$_{\odot}$ (\citealt{blair00,flan04})
and the proper-motion measurements may bear on the location of a compact
remnant such as a neutron star or black hole. After presenting our
proper-motion measurements, we return to the topic of the progenitor star
and its location in Sec.\ 4.

We measured transverse velocities of a number of optical knots and filaments
in E0102 using a software package developed by P.\ Hartigan, S.\ Heathcote,
and one of us (JAM) that is optimized to detect motions of extended nebular
objects. The software performs a subtraction-based cross-correlation following 
an algorithm described by \cite{currie}.  It has been used to measure proper
motions of shock structures and outflows in multi-epoch {\it HST} images of
the $\eta$ Carinae system and several Herbig-Haro jets
(\citealt{morse01,hart01,hart05,reip02,bally02}).
A small rectangular image region from one epoch is compared
to a larger region on the second epoch image, and the pixel-by-pixel difference
squared is calculated and summed over the region covered by the smaller image.
The sum of the squares of the differences is stored in a correlation image
for each position as the first epoch feature is raster scanned across a
section of the second epoch image.  The difference squared is a minimum where
the feature in the two epochs best aligns. After the raster scan, we invert
the correlation image so that the difference-squared minimum is a maximum.
In general, the correlation peaks (inverted minima) resembled stellar
images. We use a modified form of the flux-weighted centroiding
algorithm and uncertainty estimate in the APPHOT package of IRAF to
calculate the peak position to within a fraction of a pixel. The code
computes the mean and mean error of the x and y marginal distributions using
points within the specified centering box. 
The pixel shift of the peak from the center of the correlation
image then indicates both the magnitude and direction
of the proper motion of the feature of interest between the two epochs.
In addition to the centroiding uncertainties, there are also
systematic uncertainties due to the accuracy of the image registration
and possible photometric variations of the measured filaments. We noted
the magnitude of the former in Sec.\ 2.2; the latter could arise either
from bandpass differences of the two \HST\ cameras used, rapid radiative
cooling and evolution of the emitting filaments, or a combination of both.
Finally, by defining a ``source'' position (in this case the expansion
center of the SNR), along with the distance to the target (59 kpc) and
the time interval between observations ($\sim$8.5 years), the software
calculates the kinematic age of each feature assuming no deceleration
of the ejecta.

Regions used for measuring proper motions were selected by blinking the
registered [O~III] images on the computer screen. The difference image
in Figure~5 provides a guide to those features around the periphery of
the SNR where features appear in both epochs and outward motions are
observed. We defined individual regions around filaments and knots that
were large enough to generate clean cross-correlation peaks but small
enough to minimize differential motions within the region due to the curvature
of the outer rim of the SNR. Background stars were located in close
proximity to some of the selected ejecta filaments and had to be removed.
If stars are included in the measurement, they can skew the results
because the stars do not move in the registered images while the filaments
do, producing multiple correlation peaks. We eliminated the stars by
subtracting registered, line-free continuum images from the [O~III] images,
leaving only the emitting filaments and knots. For the WFPC2 data,
we scaled and subtracted an F547M image; for the ACS~WFC data, we scaled
and subtracted the F550M image. We avoided a few regions where subtracting
bright stars projected on or near filaments left noticeable residual flux.
Ultimately we selected 12 regions, mostly around the eastern
rim of the SNR, for measuring proper motions. These regions are identified
on the continuum-subtracted ACS~WFC [O~III] image in Figure~8.

Table~2 lists the results for our proper-motion measurements for the
12 regions we selected, where each region is labeled in Figure~8.
The table includes the region number; X and Y
pixel location in the registered images of each region center; the
tangential velocity and derived uncertainty from the centroiding
procedure; the uncertainty as a fraction of the
measured velocity; the position angle of the region center relative to
the chosen expansion center (PA-Center), the position angle of the direction
of motion (PA-Motion), and the difference between these two angles
($\Delta$PA); and the inferred kinematic age. The bottom row shows average
values of some of the measured quantities.  In particular, we measured
tangential velocities ranging from $\sim$1350 -- 2940 \kms, and an
average expansion of 1966 $\pm$ 193 \kms. This average expansion velocity
corresponds to an offset of $0.060'' \pm 0.006''$, or slightly over one
pixel, during the 8.5 year time baseline. 
The uncertainties listed in Table~2 reflect
the quality of the correlation measurements, and are approximately
commensurate with the r.m.s.\ uncertainty in the image alignment.
We must also recognize that photometric changes can occur in the
rapidly cooling, metal-rich optical filaments as the reverse shock
propagates through the ejecta (e.g., see the discussion
in \citealt{morse04}) that could affect the proper-motion measurements,
though similar debris in Cas~A tend to remain visible for several
decades (\citealt{thor01}).

Calculating the kinematic age requires that we supply an estimate of
the expansion center, i.e., where the explosion occurred. Table~2
assumes this to be the geometric center of the SNR, estimated from the
ring of reverse shock emission in the
{\it Chandra} X-ray image.  Using the IRAF routine RINGPARS in the CTIO FABRY
package, intensity peaks in horizontal and vertical cross-cuts were 
measured using Gaussian fits and the (X,Y) center of the 
X-ray ring was calculated. The geometric center is located at a position of 
RA 01\super{h}04\super{m}02\super{s}.08, DEC --72$^{\circ}$01$'$52$''$.5 
(J2000). Using this center, as marked on Figure~8 with a `+',
the average inferred kinematic age is $\sim$2123 years.

Alternatively, we empirically determined the expansion center by assuming
that the transverse motions of the ejecta are radially outward. Figure~8
shows the directions of the proper motions for all the regions we
measured. Assuming radial motions, these arrows should, in principle,
intersect at the position of the explosion. Even with the uncertainties
in the individual measurements, it is apparent when looking at Figure~8 that
the arrows on average trace back to a position south of the X-ray geometric
center. Using vector addition --- i.e., forcing the average $\Delta$PA to
be zero degrees --- we derive a proper-motion center at
RA 01\super{h}04\super{m}02\super{s}.05, DEC --72$^{\circ}$01$'$54$''$.9 (J2000).
Using this expansion center position, Table~3 lists the new position angles for
each region and a revised average kinematic age of 2054 $\pm$ 584
years.\footnote[12]{Throwing out the highest and lowest inferred ages yields
the same average age and a reduced uncertainty of 2053 $\pm 450$ years.
Throwing out the two highest and two lowest values also yields
a similar age and further reduced uncertainty of 2033 $\pm 320$ years.}
The 1$\sigma$ uncertainty in the average $\Delta$PA is $\pm$14.8 degrees.
Projecting this back to the derived expansion center position, we define
a 1$\sigma$ error circle of radius $3.4''$. Our derived expansion center
(marked by an `X') and 1$\sigma$ error circle are drawn in Figure~8.
The geometric center of the X-ray ring is located slightly east and $\sim2.4''$
to the north ($\sim0.7\sigma$ in projection) of the derived expansion center.

\section{Discussion} \label{disc}

\subsection{Geometry, Expansion Center, and Kinematic Age}

The ability to see the full velocity range of optical [O~III] emission
in E0102 (Figure~3) adds to our understanding of the geometry and dynamics
of this SNR. Because much of the emission from the central region seen in
the ACS data was Doppler-shifted out of the narrow bandpass of the WFPC2 data,
we conclude that these high-velocity ejecta must lie on the front or rear side
of an expanding shell. The [O~III] Fabry-Perot data cube of \cite{erik01}
shows that most of this emission is blueshifted up to $\sim-2100$ \kms.
Neither this highly blueshifted emission nor several other filaments through
the middle of the SNR have X-ray kinematic counterparts while in many other parts
around the outer rim the association seems clear. The optical emission
distribution requires that a shell-like component be included in models
of E0102, even if most of the mass traced in X-rays is confined to a
ring-like structure (\citealt{flan04}). For completeness we note that,
to our eyes, the main ring of (reverse shock) X-ray emission in the
{\it Chandra} images shown in \cite{gaetz00}, \cite{flan04}, and our
Figures~6 and 7 is clearly not circular, as has usually been assumed
in previous models, but is somewhat elliptical with the major axis
aligned along PA $\sim -15^{\circ}$.

Our proper-motion results provide vital missing pieces to the E0102 puzzle
and allow us to pull together a number of loose threads. 
Our derived expansion center projects in the middle of the backward
S-shaped filament complex we described above and aligns with the break in the
optical rim in the southwest that is reminiscent of the Cas~A northeastern jet.
\cite{erik01} noted peculiar [O~III] kinematics in the central region of their
Fabry-Perot data cube, i.e., low-velocity emission projected against the fast-moving
approaching debris (see also the discussion in Flanagan et al.\ 2004). Moreover,
our result places the expansion center at or near to the X-ray hotspot in
the SNR middle to which all the X-ray spokes attach (\citealt{gaetz00,flan04}).
It also coincides with the central radio emission blob that \cite{amyball}
suspected may be the site of a plerion. In fact, a bar of radio emission,
surrounded by X-ray emission, runs northeast-southwest along PA $\sim50^{\circ}$
directly through our expansion center. It is possible that this structure could be
jet-induced, similar to Cas~A (\citealt{hwang04}), and that both the
radio and X-ray blobs trace emission associated with a compact object,
perhaps a neutron star or black hole. 

Our proper-motion age for E0102 of 2054 $\pm$ 584 years agrees well with
the expansion age of $\sim$2100 years estimated by \cite{erik01} based on their
fitting an expansion ellipse to Fabry-Perot [O~III] radial-velocity data, 
but is only marginally consistent with the X-ray expansion age 
of $\sim$1000 years estimated by Hughes et al. (2000).
We place the site of the explosion squarely at the 
X-ray/radio blob of interest just south of the geometric center. 
The asymmetric distribution of optical ejecta and the slightly oval-shaped
morphology in the X-rays could be due to an asymmetric explosion or
possibly a density gradient in the circumstellar medium
(mass loss from the progenitor star). For the latter case, we would
infer higher circumstellar densities in the south, east, and near side,
and lower densities in the north, west, and far side.
Using the {\it Chandra} image, we measure a blast wave expansion radius
from our derived expansion center of 7.2 pc ($\sim25.2''$) towards the north
roughly at PA $\sim -15^{\circ}$ and 5.6 pc ($\sim19.5''$) towards the south at
PA $\sim165^{\circ}$. The expansion has proceeded $\sim$28\% farther to the 
north than to the south. Adopting an expansion age of 2100 years, the blast wave
has propagated to the north with a mean velocity of $\sim3360$ \kms\ and to
the south with a mean velocity of $\sim2610$ \kms. Probably the blast wave
began expanding at a speed of $\gtsim10,000$ \kms\ and has now slowed (as
circumstellar material is swept up) to a value somewhat below the implied
mean velocity in a particular direction.
We predict not only a north-south asymmetry in a
future X-ray measurement of the blast wave expansion, but also higher reverse
shock velocities and lower densities in the north, consistent with the lack of
optical emission. Higher X-ray temperatures to the north and higher densities
to the south are precisely what \cite{sasaki} infer in a new analysis
of a combined far-UV and X-ray data set on E0102. Because of the noted
asymmetries, we urge caution in
using global average values to describe the geometry, dynamics, and evolution
of E0102 when comparing its parameters to model predictions (e.g., \citealt{chev05});
both the X-ray and optical data imply directional variations.

\subsection{Clues to the E0102 Progenitor: Spying on the Neighbors}

The E0102 progenitor may have been a very massive main-sequence star that
underwent substantial mass loss during its late stages of evolution. 
The resulting SN explosion was likely of Type~IIL/b or Ib/c (\citealt{blair00,chev05}),
depending on whether a thin H envelope was present when the explosion occurred (to date
there is no indication of fast-moving H or He debris mixed in
with the expanding ejecta). The fast-moving O-rich ejecta we observe
are broadly consistent with the predictions of models of nucleosynthesis
within the helium-burnt layers of evolved massive stars
(\citealt{woosley,thiel,nomoto}), though in detail comparing to any
particular set of models is complicated by an object's specific initial
conditions (i.e., metallicity, ZAMS mass, binarity)
and by the fact that only some of the ejecta are emitting and can be analyzed.

\cite{blair00} compared their relative elemental abundance estimates to the
nucleosynthesis predictions of \cite{nomoto} and found the best agreement
with models that had progenitor ZAMS masses $>$ 25M$_{\odot}$. \cite{flan04}
obtained similar relative abundance ratios from the {\it Chandra} data,
especially for O/Ne, but also estimated a total oxygen mass in the ejecta
of $\sim$6 M$_{\odot}$. Again comparing to the Nomoto et al.\ models, they
inferred a progenitor ZAMS mass of $\sim$32 M$_{\odot}$. Both Blair et al.\ and
Flanagan et al.\ noted that the models under-predict the amount of Ne by
a factor of $\sim$2. The models also predict that O-burning products such as
S, Ca, and Ar should be observed, but in fact they are not, leading us to
question the true relevance of these models for making detailed comparisons.

In order to guide theoretical efforts to model the progenitor and SN type
of E0102, we suggest that meaningful constraints can be gleaned by examining
the local stellar population with which E0102 is associated. Such constraints
will be independent of the ejecta abundances and masses --- as well as their
assumptions about geometries, ionization states, filling factors, etc.\ ---
inferred from observations of E0102 itself. For example, one of the reasons for
believing that $\eta$ Carinae is an evolved supermassive star is that there are
main-sequence O3 stars in close proximity (e.g., \citealt{walborn}). One needs
to be wary of issues such as actual physical co-location and coeval formation,
but for E0102 there appears to be a relevant connection.

Figure~1 shows the N76A nebula that surrounds the OB association Hodge 53
(cf., \citealt{massey00}). E0102 lies a projected distance of only $\sim$10 pc
to the northeast of the nebula edge. N76A's high ionization requires a high 
effective temperature for the main exciting source SMC~WR7
(= AzV~336a; \citealt{pakull91}). Moreover the round appearance of N76A
is more consistent with a Wolf-Rayet bubble than a star-forming H~II region
like the Orion nebula. \cite{niemela02} found that SMC~WR7 is a double-lined
WN2 + O6 spectroscopic binary with a period of 19.56 d. These authors constrain
the minimum masses to be $\sim$18 and 34 M$_{\odot}$ for the WN2 and O6 components,
respectively. In addition, although coeval formation is questionable, \cite{massey00}
suggest that the ZAMS mass of the WN2 progenitor star in the
SMC~WR7 system is $>50$ M$_{\odot}$ based upon the turn-off mass of the cluster.
We speculate that a plausible progenitor candidate for E0102 is a star 
with properties similar to the WN2 component of SMC~WR7 (\citealt{massey01}), 
given the similar strong He~II $\lambda4686$ emission
of the E0102 halo and the general agreement of the fast-moving
ejecta abundances with models of Wolf-Rayet
atmospheres (e.g., \citealt{woosley}). For such a progenitor star, the
SN explosion was likely of Type Ib/c.

\section{Conclusion} \label{conc}

A new imaging data set from the {\it Hubble Space Telescope} Advanced Camera
for Surveys reveals spectacular details in the O-rich SNR 1E~0102--7219
in the SMC. We used the broad-band SDSS $g$ (F475W) filter to capture the full
[O~III]$\lambda\lambda4959,5007$ velocity range of the ejecta.
Many high-velocity knots and filaments projected through the middle of the SNR
are seen for the first time at $\sim0.1''$ resolution. We discuss their
morphology, kinematics, and relationship to soft X-ray emission observed
in a {\it Chandra} ACIS-S image. The wide field of view of the ACS images
also allows us to examine the circumstellar environment of E0102. A narrowband
H$\alpha$+[N~II] (F658N) image shows considerable inhomogeneity in the
surrounding medium and a possible physical association between some of the gas
in the halo of E0102 and the nearby N76A nebula.

We compared the 2003 ACS [O~III] image to a 1995 narrowband WFPC2 Planetary
Camera [O~III] image and measured tangential expansion velocities in 12 regions
around the periphery of the SNR. We find a mean expansion velocity for the optical
ejecta of $\sim 2000$ \kms\ and a kinematic age for E0102 of $\sim2050 \pm 580$ years.
Our kinematic age agrees with a recent estimate of \cite{erik01} based on
spatially resolved [O~III] radial-velocity data, but is a factor of two longer
than found by \cite{hughes00} using multi-epoch, low-resolution X-ray images (although
it is within 2$\sigma$). 

The ejecta proper motions we measured trace back to an expansion center
$\sim2.4''$ south of the geometric center of the {\it Chandra} X-ray image.
The 1$\sigma$ error circle of our expansion center encompasses the 
conspicuous radio and X-ray blobs noted in previous studies
(e.g., \citealt{amyball,gaetz00,flan04}). Our kinematic data provide additional
evidence for believing that a compact remnant (neutron star or black hole)
may be associated with the central hotspot emissions. E0102's proximity to the N76A 
Wolf-Rayet nebula and Hodge 53 OB association leads us to speculate that the
E0102 progenitor may have been a very massive ($>50$ M$_{\odot}$) main-sequence
star that underwent considerable mass loss as it evolved through a WN phase
before exploding as a Type Ib/c supernova.

We thank the referee for carefully reviewing the manuscript and recommending
changes that significantly improved the fidelity of the results.
This  work  has  been  supported  by  NASA  grant  NAG5-12279  to  the
University of Colorado and the COS Science Team.



\clearpage

\begin{deluxetable}{cccccc}
\tablecaption{ACS Observations of 1E0102-9217 \label{thetable}}
\tablewidth{0pt}
\tablehead{
\colhead{Data Set} & \colhead{Exposure Time} & \colhead{Date} & \colhead{Central Wavelength} & \colhead{$\Delta\lambda$} & \colhead{Filter}\\
\colhead{} & \colhead{(sec)} & \colhead{} & \colhead{(\ang)} & \colhead{(\ang)} & \colhead{}
}
\startdata
J8R802010&760&15 Oct 2003&4760&1458&F475W\\
J8R802020&900&15 Oct 2003&5580&547&F550M\\
J8R802030&720&15 Oct 2003&6584&78&F658N\\
J8R802040&720&15 Oct 2003&9445$^*$&1229&F850LP\\
J8R802050&720&15 Oct 2003&7764&1528&F775W\\

\enddata

\tablecomments{$^*$ The 850LP profile is asymmetrical, with the peak throughput
at $\sim$9000\AA.}

\end{deluxetable}

\clearpage

\begin{deluxetable}{cccccccccc}
\tabletypesize{\tiny}
\tablecaption{Proper Motions of 1E~0102-7219 and Ages Derived from X-ray Geometric Center \label{datatable1}}
\tablewidth{0pt}
\tablehead{
\colhead{Region}&\colhead{x}&\colhead{y}&\colhead{Velocity}&\colhead{V error}&\colhead{Verr/V}&\colhead{PA-Center}&\colhead{PA-Motion}&\colhead{$\Delta$ PA}&\colhead{Inferred Age}\\
\colhead{ }&\colhead{(pixels)}&\colhead{(pixels)}&\colhead{(km/s)}&\colhead{(km/s)}&\colhead{ }&\colhead{(Deg from N)}&\colhead{(Deg from N)}&\colhead{(Deg from N)}&\colhead{(Years)}\\
}
\startdata

1&817.0&1230.0&1746.8&128.4&0.07&47.5&49.0&-1.5&1840.4\\
2&733.0&1178.0&2483.6&285.4&0.12&66.4&63.9&2.5&1514.1\\
3&698.0&1041.0&2074.5&165.5&0.08&93.6&116.4&-22.8&1880.8\\
4&727.0&967.0&2588.5&284.1&0.11&108.7&94.5&14.2&1437.0\\
5&748.0&910.0&1690.7&184.5&0.11&120.6&95.0&25.6&2235.8\\
6&715.0&870.0&1356.2&135.0&0.10&123.5&120.0&3.5&3251.7\\
7&793.0&857.0&2009.5&138.2&0.07&134.2&116.6&17.6&1858.2\\
8&844.0&790.0&1392.2&80.0&0.06&149.7&141.5&8.2&2879.7\\
9&1096.0&688.0&1737.4&78.1&0.05&194.1&181.6&12.5&2828.5\\
10&1284.0&911.0&2013.0&132.5&0.07&242.1&227.7&14.4&2027.6\\
11&1250.0&970.0&2940.0&587.7&0.20&250.0&272.0&-22.0&1148.1\\
12&1226.0&1279.0&1556.9&114.2&0.07&314.3&325.7&-11.4&2572.1\\
\hline
Average& & &1965.8&192.8&0.09& & &3.4 $\pm$ 15.4&2122.8 $\pm$ 644.4\\
\enddata

\tablecomments{The first three columns give the number of each of the twelve regions,
and the pixel coordinates of those regions on the registered images.  Column four gives the
tangential velocity of each region, which was computed using our proper motion code.
The next column estimates the uncertainty in that velocity, as calculated by the centroiding algorithim
in our proper motion code.  The seventh and eight columns list two position angles (PAs), where
PA-Center is the angle from North each filament is moving with respect to the geometric center,
and PA-Motion is the angle from North each filament is moving with respect to its current position.
$\Delta$PA is the difference in these two angles.  The last column gives the inferred kinematic
age, based on the velocity given by the proper motions, the distance to the SMC, and the
time between epochs.  The geometric center, based on the X-ray emission, was assumed to be
at pixel (1002,1060), corresponding to
RA 01\super{h}04\super{m}02\super{s}.08, DEC --72$^{\circ}$01$'$52$''$.5 (J2000).}

\end{deluxetable}

\begin{deluxetable}{ccccc}
\tabletypesize{\tiny}
\tablecaption{Ages Derived from Proper-Motion Derived Center \label{datatable2}}
\tablewidth{0pt}
\tablehead{
\colhead{Region}&\colhead{PA-Center}&\colhead{PA-Motion}&\colhead{$\Delta$ PA}&\colhead{Inferred Age}\\
\colhead{ }&\colhead{(Deg from N)}&\colhead{(Deg from N)}&\colhead{(Deg from N)}&\colhead{(Years)}\\
}
\startdata

1&40.4&49.0&-8.6&2138.5\\
2&58.0&63.9&-5.9&1658.5\\
3&83.8&116.4&-32.6&1911.4\\
4&98.3&94.5&3.8&1394.1\\
5&110.7&95.0&15.7&2088.2\\
6&115.3&120.0&-4.7&3037.8\\
7&125.3&116.6&8.7&1661.5\\
8&143.3&141.5&1.8&2494.0\\
9&195.7&181.6&14.1&2446.0\\
10&250.8&227.7&23.1&1871.2\\
11&261.2&272.0&-10.8&1074.8\\
12&321.0&325.7&-4.7&2872.1\\
\hline
Average& & &0.0 $\pm$ 14.8&2054.0 $\pm$ 584.0\\
\enddata

\tablecomments{Expansion velocities and ages based on minimizing the value of $\Delta$PA.
Columns two and three list the modified PAs that were measured from minimizing $\Delta$PA.
Column four lists the $\Delta$PAs, and the last column lists the inferred kinematic age.
The proper-motion derived center lies at
RA 01\super{h}04\super{m}02\super{s}.05, DEC --72$^{\circ}$01$'$54$''$.9 (J2000).}

\end{deluxetable}

\begin{figure}
\epsscale{0.8}
\plotone{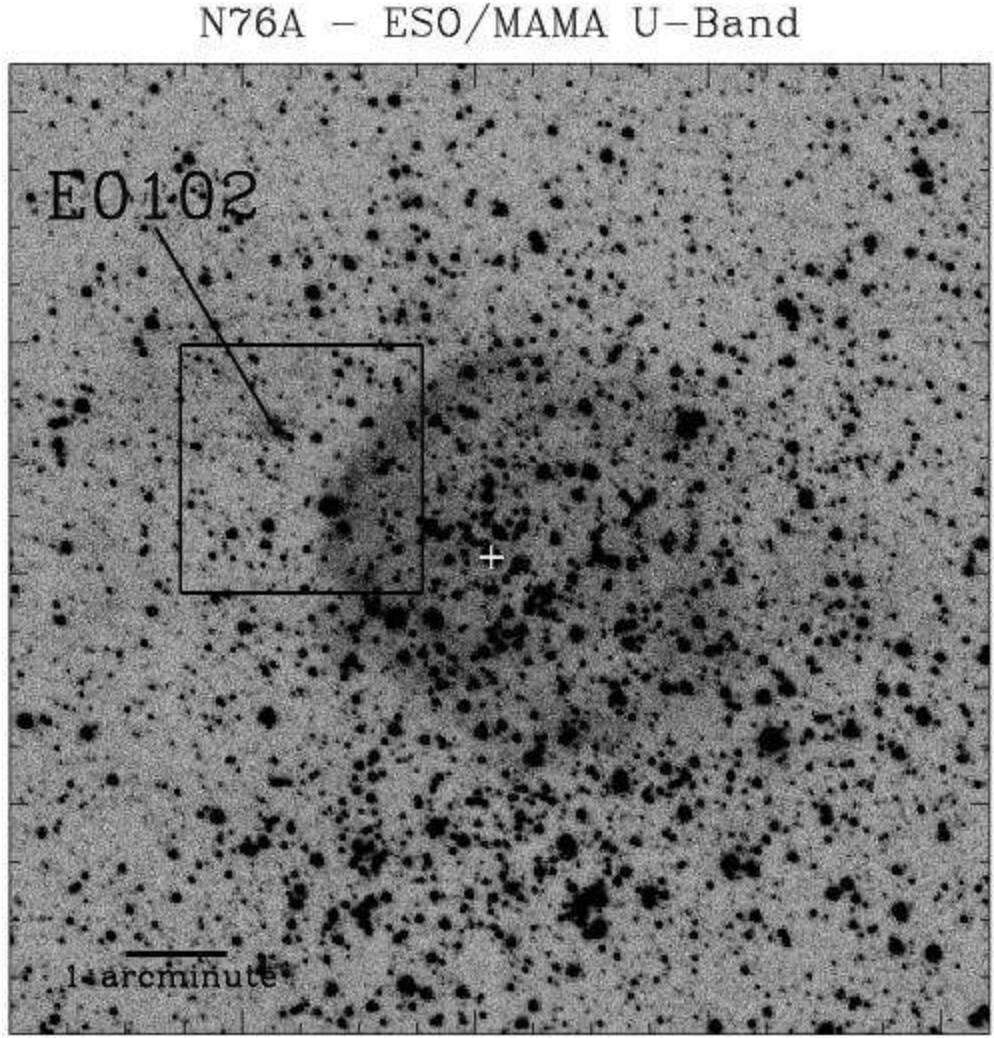}
\caption{An ESO/MAMA {\it U}-Band image of the N76A nebula and it's vicinity. 
This image includes [OII]$\lambda$$\lambda$3727 emission from the N76A nebula and
the SNR E0102.  The box shows the ACS~WFC footprint of the field shown in Figs.\ 2 and 3,
and the white plus sign marks the position of SMC~WR7, the WN2 + O6  binary system that
ionizes the nebula.}
\end{figure}

\begin{figure}
\epsscale{0.8}
\plotone{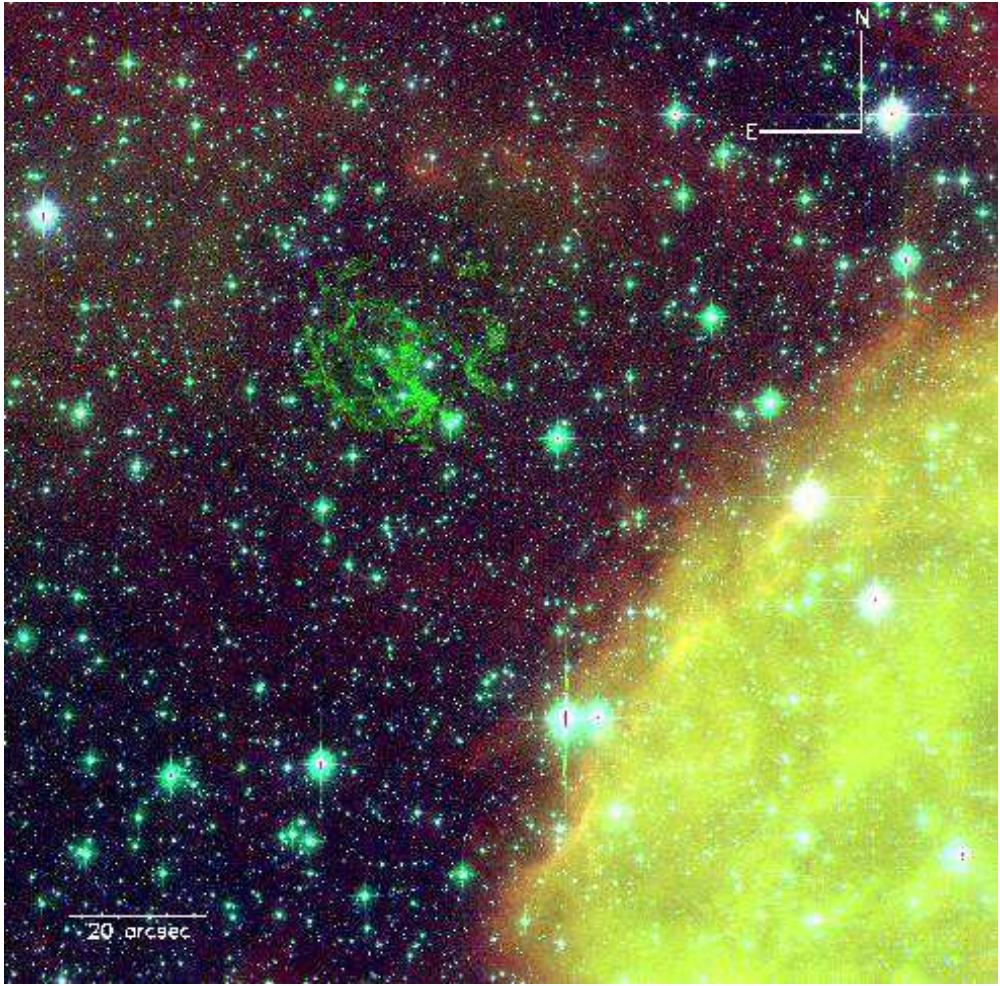}
\caption{A three color ACS~WFC image of E0102 and its  surroundings,
with H$\alpha$ + [NII] F658N shown in  red, [O~III]  F475W shown  in green,  and the
continuum F550M shown  in blue.  E0102 appears green because it is
O-rich and  lacking in  hydrogen.  There is  a cleared  region between
E0102 and the surrounding halo that shows the progress of the
forward  shock; this  region is  filled with X-rays.   The nearby
high-excitation N76A nebula is in the  lower-right corner.}
\end{figure}

\begin{figure}
\epsscale{.5}
\plotone{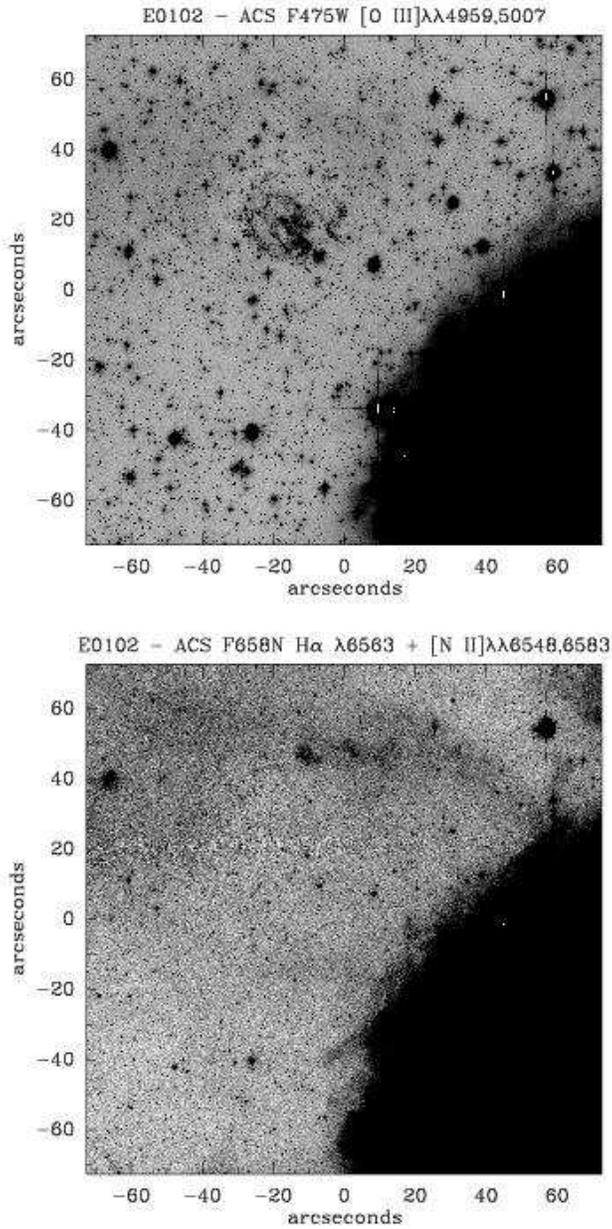}
\caption{({\it a}) [O~III] emission from  the F475W
filter.  ({\it b}) H$\alpha$ + [NII] emission  from the F658N
filter.  Due to its lack  of hydrogen, E0102 is completely absent from
the bottom H$\alpha$ + [NII] image.  The (0,0) position of these  images 
is at  RA 01\super{h}03\super{m}58\super{s}.7, DEC --72$^{\circ}$02$'$14$''$.8
(J2000).}\label{ohcomp}
\end{figure}

\begin{figure}
\epsscale{1.0}
\plotone{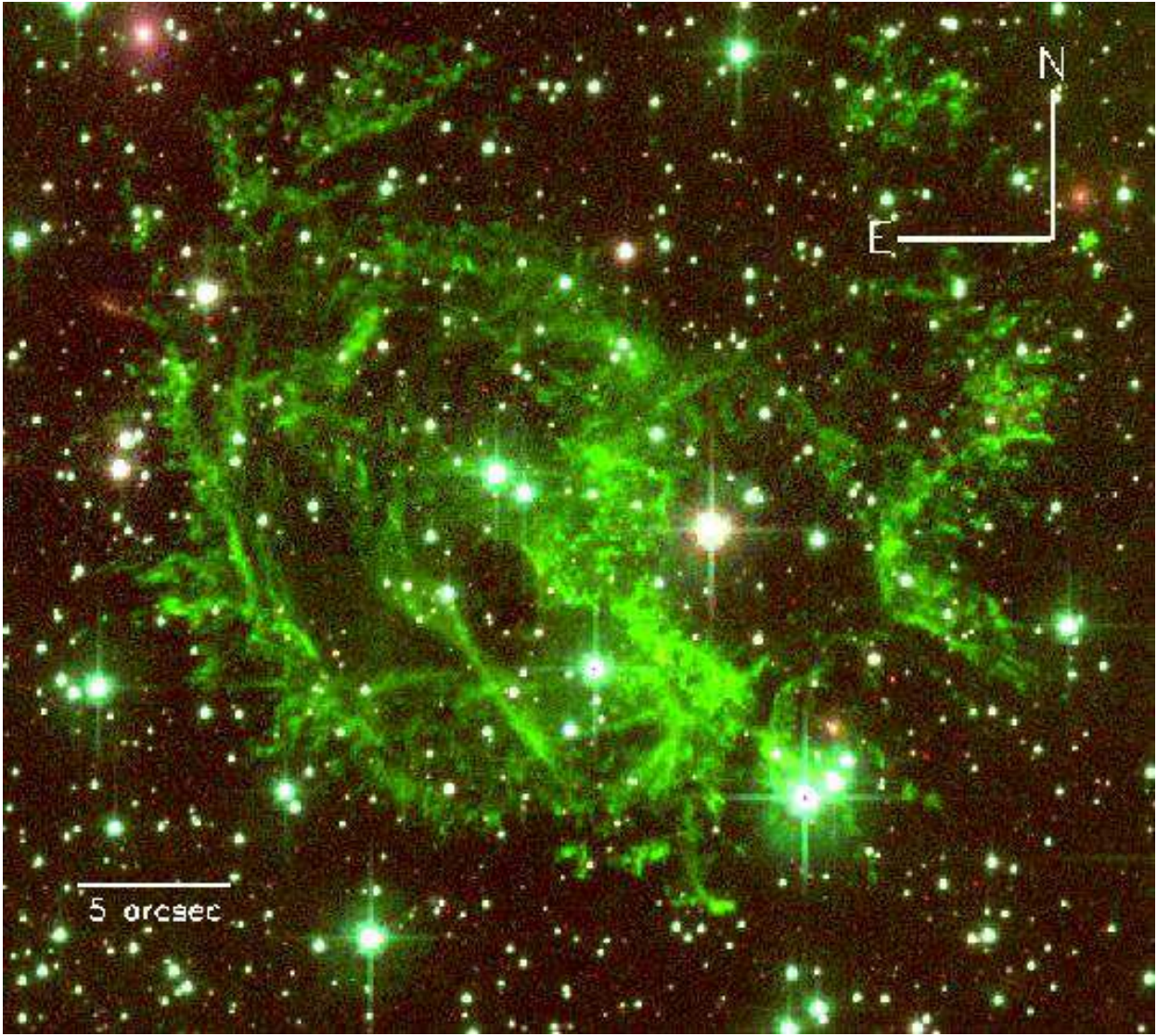}
\caption{A  three  color  ACS~WFC close-up image  of E0102.   Green  represents  [O~III]
emission from  the F475W filter,  blue represents the  continuum F550M
filter,  and  red represents  the  F775W  filter.   The full  velocity
structure of the  SNR is seen here, showing much  of the emission that
was Doppler shifted out of the  WFPC2 data.  The backward S-shaped feature that
runs from the northeast to the south-southwest is likely associated with the approaching
side of the ejecta shell.  The optical distribution is asymmetrical with most of the
emission in the southeast quadrant.}\label{color2}
\end{figure}

\begin{figure}
\epsscale{0.9}
\plotone{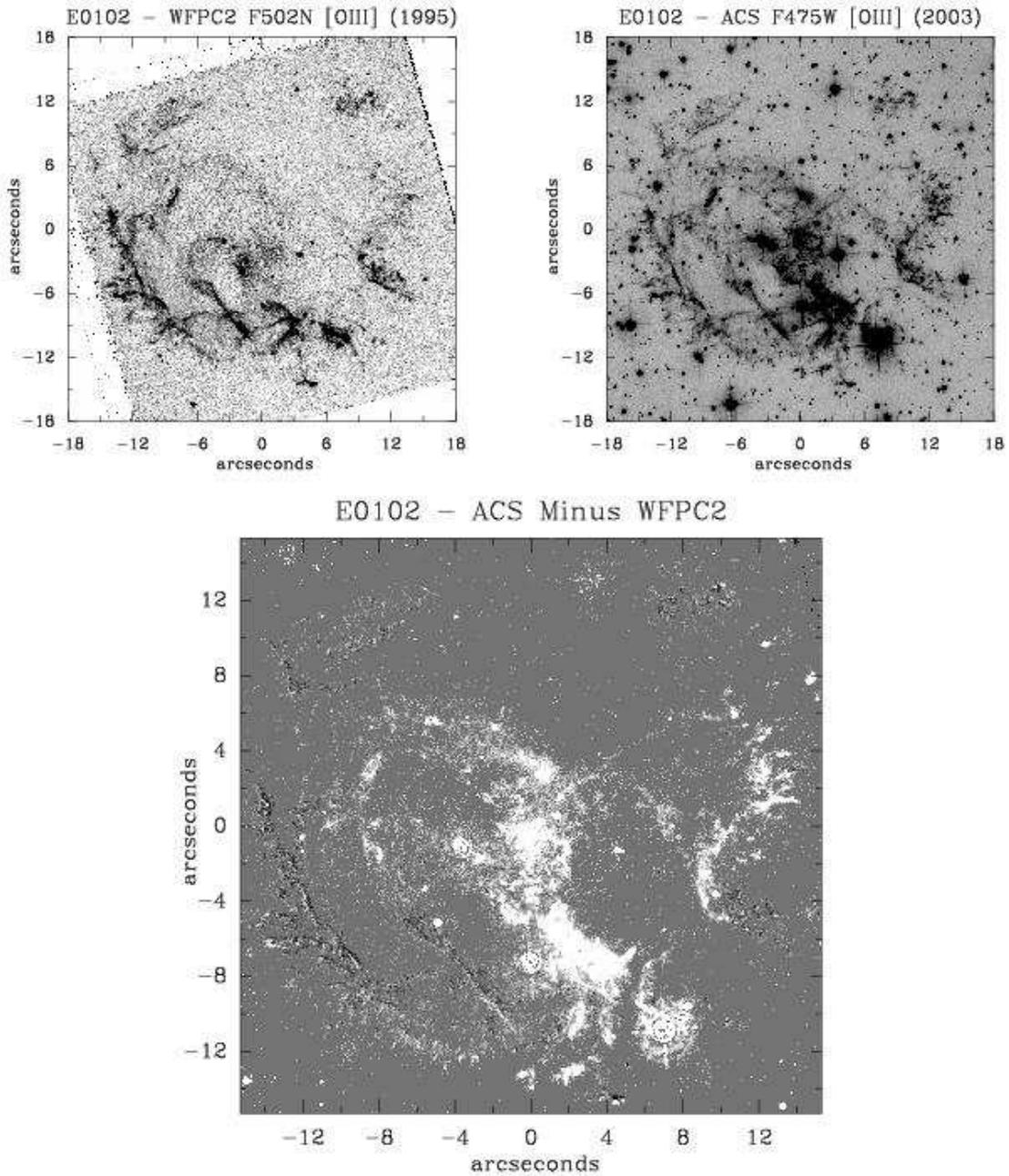}
\caption{({\it Top}) WFPC2  1995 epoch and  ACS 2003  epoch [O~III] images  of 1E~0102--7219.
These images have been registered using background stars
(see Figure 6).  The additional emission and increased signal-to-noise are
readily apparent in the ACS image. ({\it Bottom}) A difference image highlighting the motion
detected between the two epochs, as well as new emission detected in the 2003 epoch.
The ACS 2003 epoch  appears white, while the WFPC2 1995 epoch appears black.}\label{twoims}
\end{figure}

\begin{figure}
\epsscale{1.0}
\plottwo{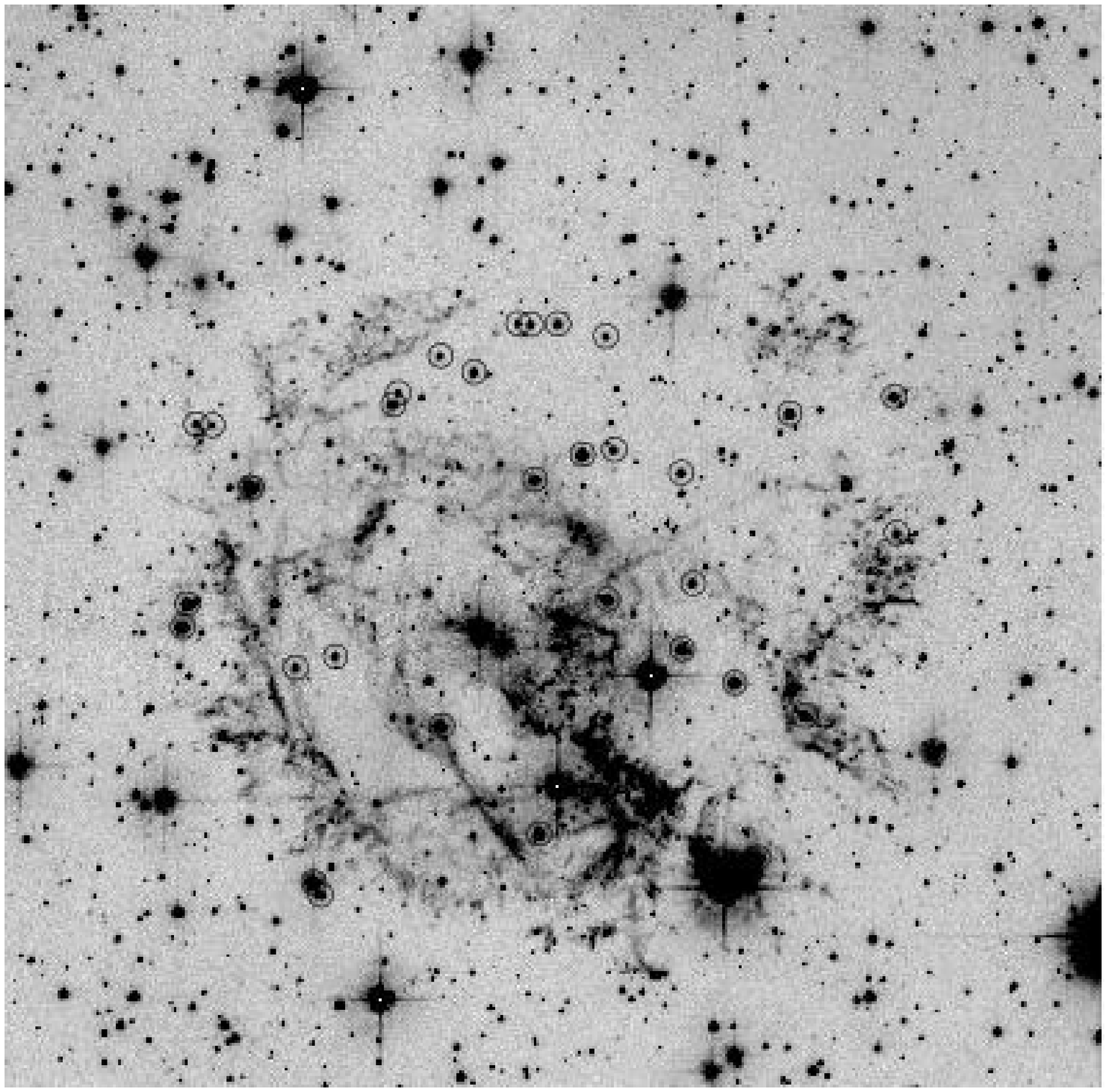}{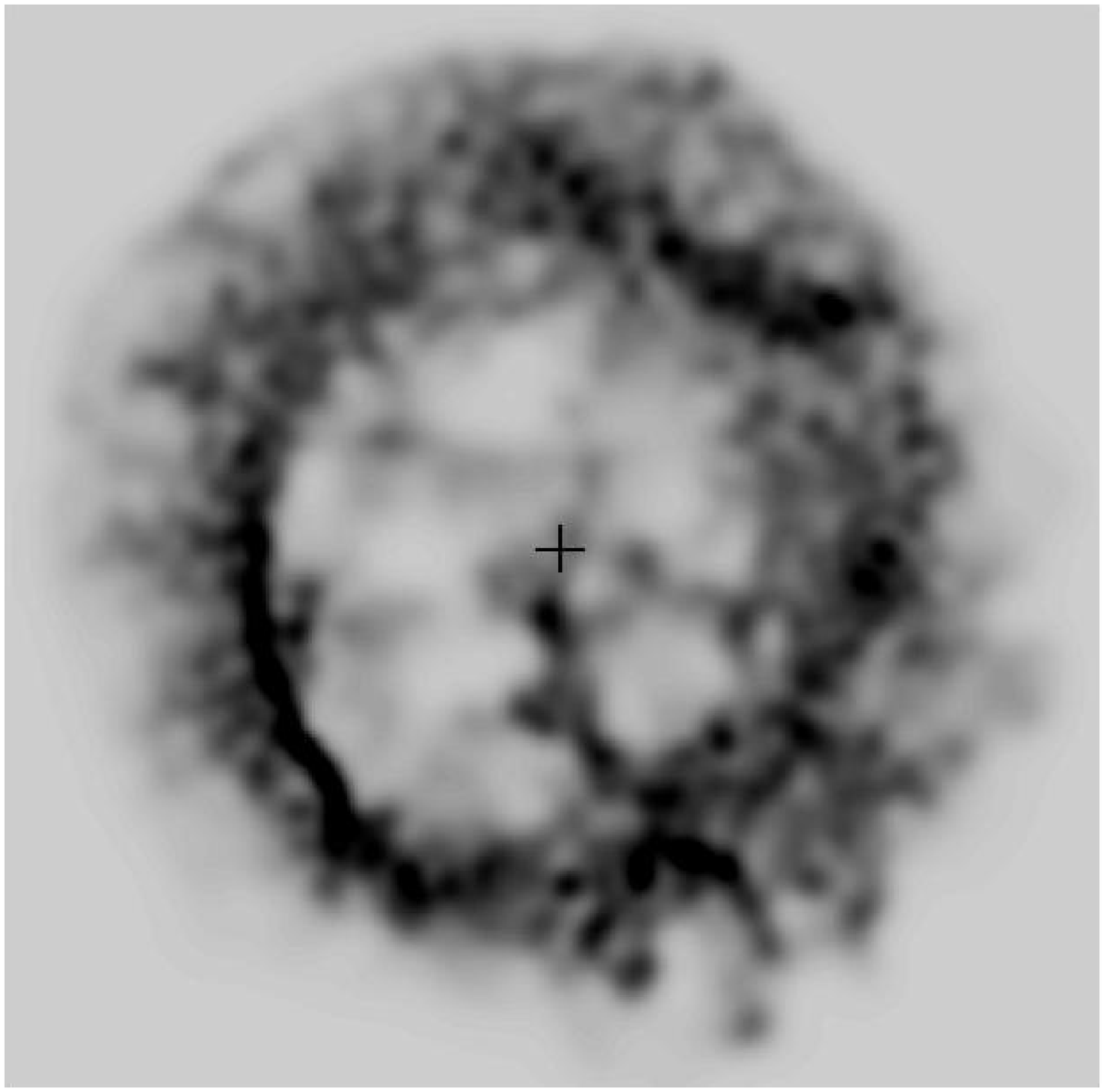}
\caption{({\it Left}) ACS [O~III] 2003 epoch  image of 1E~0102--7219.  Circles are
drawn around the 40 stars that  were used to register  this image to
the WFPC2 F502N  1995 epoch image (some of the stars are hidden by emission).  ({\it Right}) {\it Chandra} ACIS-S 1999 X-ray
image of E0102.   We calculated the geometric center using horizontal and vertical
intensity crosscuts.  This center is denoted by the plus sign in this image and
in Figure 8.  The field shown is the same in both figures.}\label{regcircles}
\end{figure}

\begin{figure}
\epsscale{1.0}
\plottwo{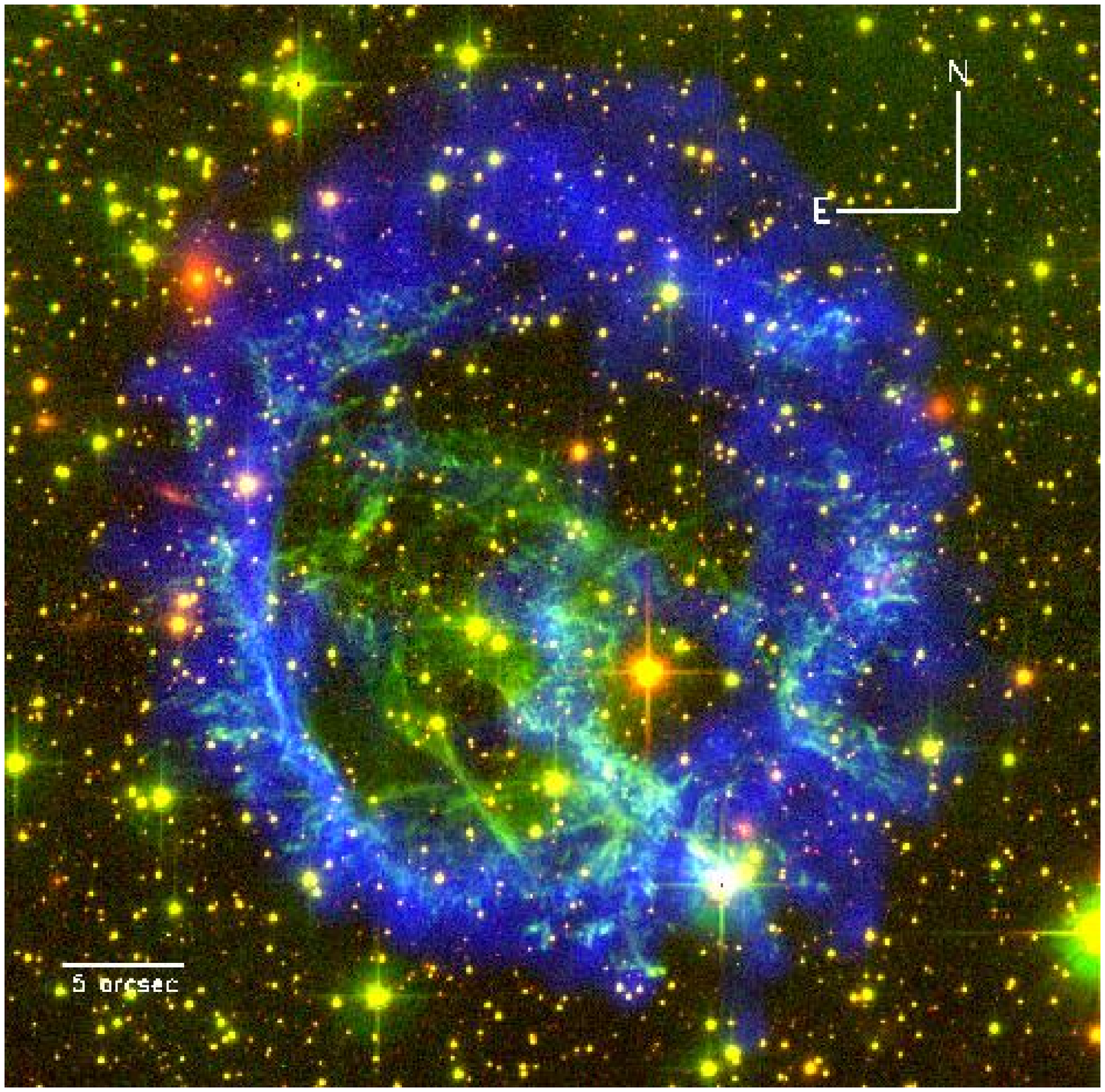}{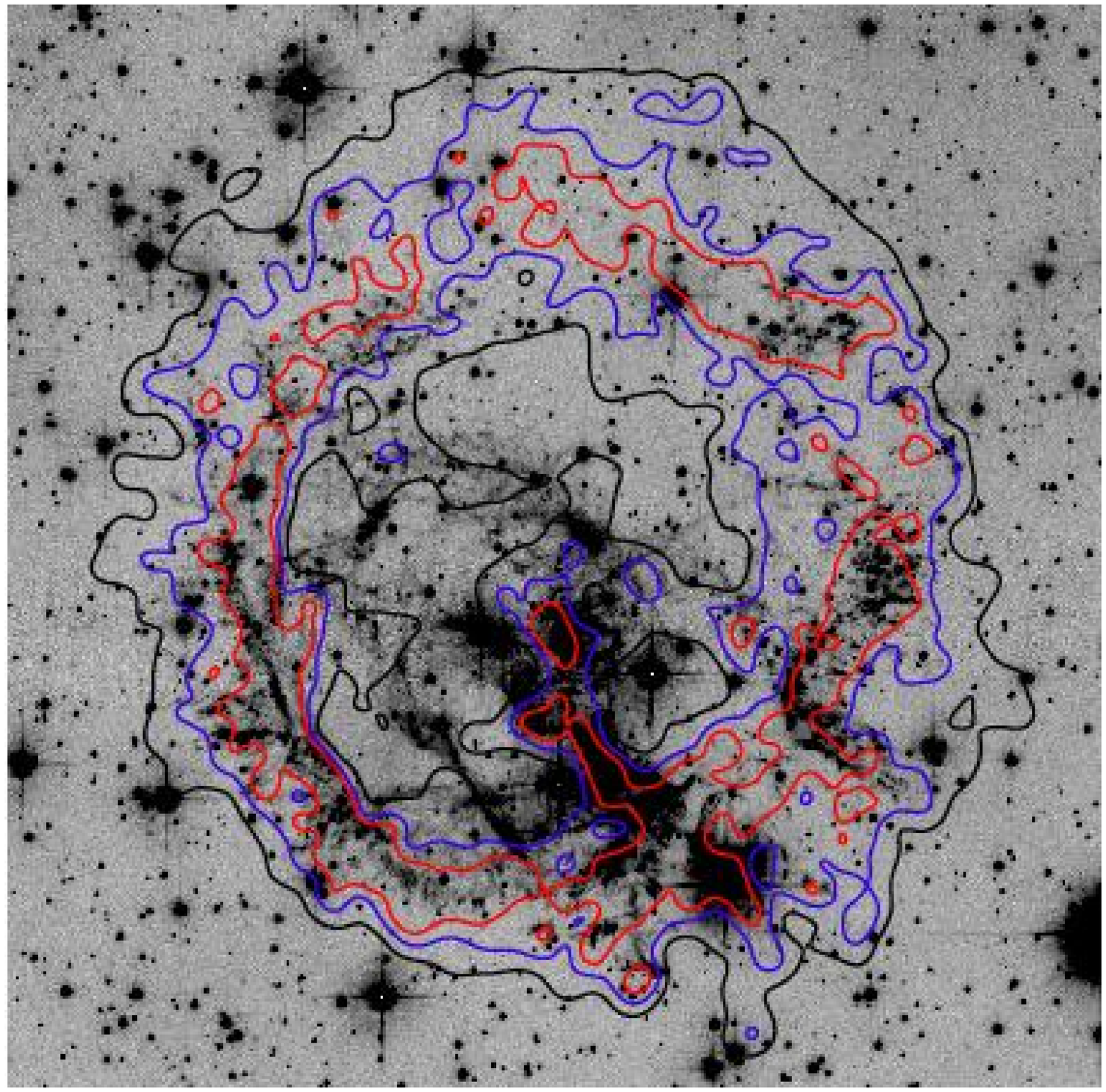}
\caption{({\it Left}) Three color image, with the  ACS 2003 [O~III] in green and the
F775W filter  in red.   The 1999 {\it Chandra}  image is represented  in
blue. ({\it Right}) ACS 2003 epoch is shown with contours from the
1999  {\it Chandra}  image   overlaid.  The red contour marks the brightest X-ray emission 
and is coincident with the reverse shock, while the  black  contour outlines the faintest 
emission with the outer edge at the position of the forward  blast
wave.  The X-ray and optical ejecta emissions correspond in many areas, but are also 
anti-correlated in several regions.}\label{contour}
\end{figure}

\begin{figure}
\epsscale{1.0}
\plotone{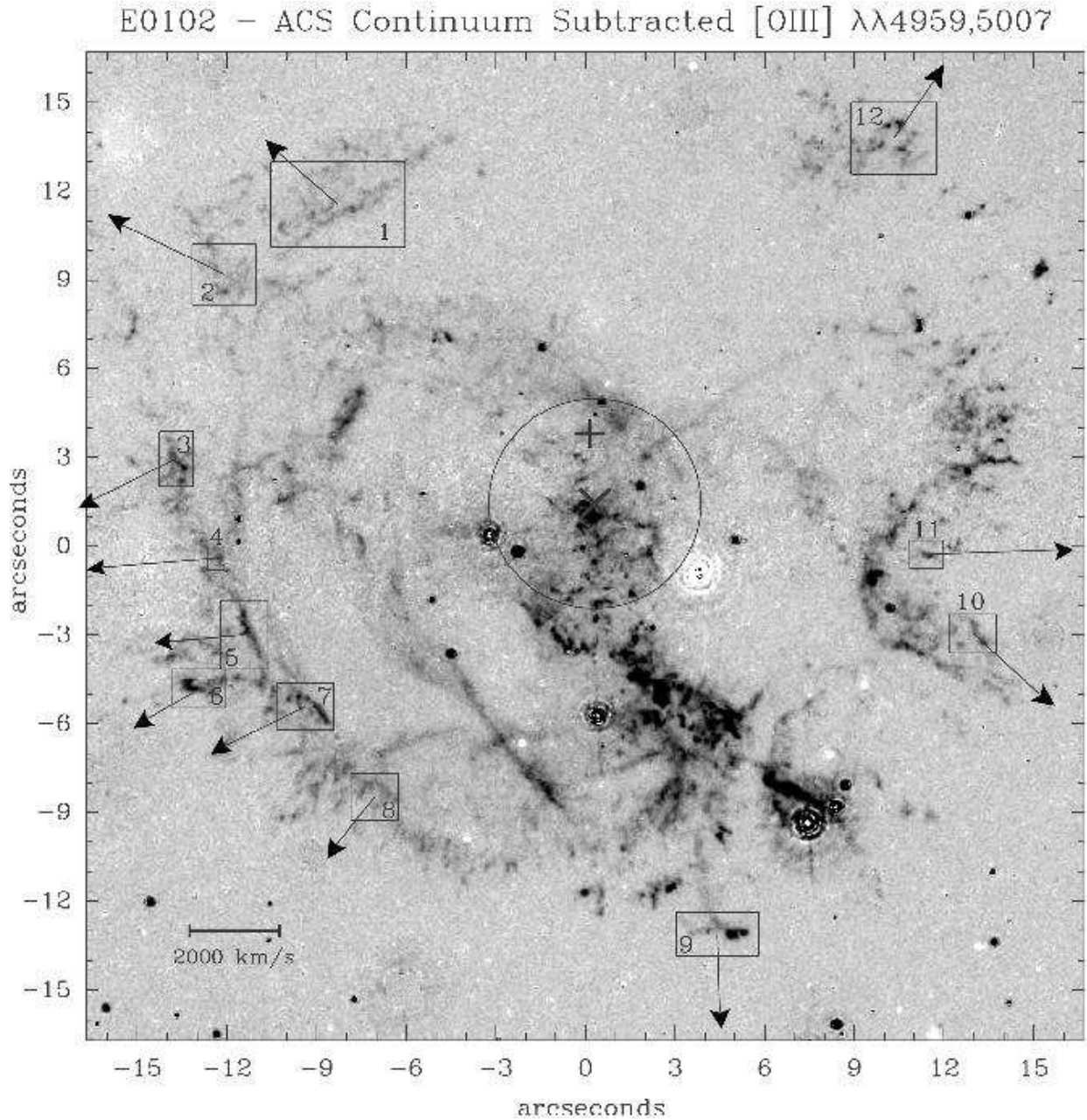}
\caption{The ACS continuum-subtracted [O~III] image from 2003  with boxes drawn around  the 
regions  that had detectable proper  motions.  The arrows show the  directions of motion
of the measured filaments.   The line lengths accurately represent the
velocity magnitudes, with $1.5''$ = 1000 \kms.  The  + marks  the
geometric  center,   and  the $\times$ marks  the   proper-motion  derived
center.  A 1$\sigma$ error circle is drawn around the proper-motion  derived center 
(see text for details).}\label{igiarr}
\end{figure}

\end{document}